\documentclass[]{egu} 
\begin{document}

\title{Spectra and anisotropy of magnetic fluctuations in the Earth's magnetosheath: Cluster observations}

\runningtitle{Magnetosheath turbulence anisotropy}  

\runningauthor{O.~Alexandrova et al.} 

\correspondence{O.~Alexandrova or C.~Lacombe\\ (alex@geo.uni-koeln.de, catherine.lacombe@obspm.fr)} 

\author{O.~Alexandrova}
\affil{University of Cologne, Institute of Geophysics and Meteorology, Albertus-Magnus-Platz 1, 50923 Cologne, Germany}
\author{C.~Lacombe}
\author{A.~Mangeney}
\affil{LESIA, Observatoire de Paris, CNRS, UPMC, Universit\'e Paris Diderot, 5 place J.~Janssen, 92190 Meudon, France}

\maketitle 

\begin{abstract}
We investigate the spectral shape, the anisotropy of the wave vector distributions and the anisotropy of the amplitudes of the magnetic fluctuations in the Earth's magnetosheath within a broad range of frequencies $[10^{-3},10]$~Hz which corresponds to spatial scales from $\sim 10$ to $10^5$~km. We present the first observations of a Kolmogorov-like inertial range of Alfv\'enic fluctuations $\delta B^{2}_{\perp}\sim f^{-5/3}$ in the magnetosheath flanks, below the ion cyclotron frequency $f_{ci}$. 
In the vicinity of $f_{ci}$, a spectral break is observed, like in solar wind turbulence. Above the break, the energy of compressive and Alfv\'enic fluctuations  generally follow a power law with a spectral index between $-3$ and $-2$. 
Concerning the anisotropy of  the  wave vector distribution, we observe a clear change in its nature in the vicinity of ion characteristic scales: if at MHD scales there is no evidence for a dominance of a slab ($k_{\|}\gg k_{\perp}$) or 2D ($k_{\perp}\gg k_{\|}$) turbulence, above the spectral break, ($f>f_{ci}$, $kc/\omega_{pi} > 1$) the 2D turbulence dominates. 
This 2D turbulence is observed in six selected one-hour intervals among which the average proton $\beta$ varies from $0.8$ to $9$.  It is observed for both the transverse and compressive magnetic fluctuations, independently on the presence of linearly unstable modes at low frequencies or Alfv\'en vortices at the spectral break. 
We then analyse the anisotropy of the magnetic fluctuations  in a time dependent reference frame based on the field ${\bf B}$ and  the flow velocity ${\bf V}$ directions. Within the range of  the 2D turbulence, at scales $[1,30]k c/\omega_{pi}$, and for any $\beta$ we find that the magnetic fluctuations at a given frequency in the plane perpendicular to ${\bf B}$ have more energy along the ${\bf B} \times {\bf V}$ direction.  This  non-gyrotropy of the fluctuations at a fixed frequency is consistent with gyrotropic fluctuations at a given wave vector, with $k_{\perp}\gg k_{\|}$, which suffer a different Doppler shift along and perpendicular to ${\bf V}$ in the plane perpendicular to  ${\bf B}$.
 \end{abstract}

\section{Introduction}
In  the space plasma turbulence, the presence of a mean magnetic field ${\bf B}$ gives rise to anisotropies with respect to the field direction (${\parallel}$ means parallel, and ${\perp}$ means perpendicular to ${\bf B}$). There are anisotropies both in the intensities $\delta B^2$ of the magnetic fluctuations ($\delta B^2_{\perp}\ne \delta B^2_{\|}$) and in the distribution of their wave vectors ${\bf k}$ ($k_{\perp}\ne k_{\|}$), i.e., the energy distribution of the turbulent fluctuations is anisotropic in ${\bf k}$--space.

To study the anisotropy of  turbulent fluctuations in space plasma, we chose here the Earth's magnetosheath as a laboratory. Downstream  of the bow shock, the solar wind plasma slows down, and the plasma density, temperature and magnetic field increase in comparison with the solar wind plasma.   The magnetosheath boundaries (bow shock and magnetopause) introduce an important temperature anisotropy $T_{\perp}>T_{\|}$, and  therefore linearly unstable waves,  such as Alfv\'en Ion Cyclotron (AIC) and mirror modes, are present (see the reviews by Schwartz et al., 1996; Lucek et al., 2005; Alexandrova, 2008).  In the vicinity of the bow-shock,  an  $f^{-1}$ power law spectrum is observed at frequencies below the ion cyclotron frequency, $f<f_{ci}$, \citep{czay01}. The power law spectra $\sim f^{-5/3}$, typical of the solar wind inertial range at $f<f_{ci}$, have not been observed in the magnetosheath.  However, as in the solar wind, the energy of  the  magnetic fluctuations follows a power law close to $\sim f^{-3}$ at frequencies $f>f_{ci}$ \citep{rezeau99,czay01}.

The question of the anisotropy of wave vectors in the magnetosheath has been mostly addressed for dominant frequencies in the turbulent spectrum (spectral peaks), below $f_{ci}$, where  linearly unstable modes are expected (Sahraoui et al. 2004; Alexandrova et al., 2004; Sch\"afer et al., 2005; Narita et al., 2006; Narita and Glassmeier, 2006; Constantinescu et al., 2007). Instead, we are interested in permanent fluctuations in the magnetosheath (and not in spectral peaks) which cover a very broad range of frequencies (more than 5 decades), from frequencies well below $f_{ci}$ to frequencies much higher than $f_{ci}$. 

These permanent fluctuations within the frequency range $[0.35,12.5]$~Hz, above $f_{ci}$, and for one decade of scale lengths around Cluster separations ($\sim 100$~km), have been studied by Sahraoui et al. (2006) using the $k$-filtering technique. For a relatively short time interval in the inner magnetosheath (close to the magnetopause) and for a proton beta $\beta_p \sim 4$, the authors show that the wave-vectors of the fluctuations are mostly perpendicular to the mean magnetic field ${\bf B}$, $k_{\perp}\gg k_{\|}$, and that their frequency $\omega_0$ in the plasma frame is zero. In the plane perpendicular to ${\bf B}$, the ${\bf k}$-distribution is non-gyrotropic, more intense and with a well-defined power law $k^{-8/3}$ in a direction along the flow velocity ${\bf V}$  which was perpendicular to both  ${\bf B}$ and the normal to the magnetopause for this particular case. The presence of linearly unstable  large scale mirror mode during the considered time interval makes the authors   conclude  that the small scale fluctuations with the observed dispersion properties $k_{\perp}\gg k_{\|}$ and $\omega_0 = 0$ result from  a non-linear cascade of mirror modes.

At higher frequencies, $\sim[10,10^3]$~Hz, between about  the lower hybrid frequency $f_{lh}$ and 10 times the electron cyclotron frequencie $f_{ce}$, the permanent fluctuations observed in the magnetosheath, during four intervals of several hours, have been studied by Mangeney et al. (2006) and Lacombe et al. (2006). The corresponding spatial scales, $\sim [0.1,10]$~km $\simeq [0.3,30] kc/\omega_{pe}$ ($c/\omega_{pe}$ being the electron inertial length), are much smaller  than the Cluster separations, and so only  the one-spacecraft technique could be used to analyze the anisotropy of wave vector distributions.

Magnetic fluctuations with $k_{\|}\gg k_{\perp}$,  usually called {\it slab turbulence}, have rapid variations of the correlation function along the field and weak dependence upon the perpendicular coordinates. For the fluctuations with $k_{\perp}\gg k_{\|}$, called {\it 2D turbulence}, the correlation function varies rapidly in the perpendicular plane, and there is no dependence along the field direction. So, measurements along different directions with respect to the mean field can give the information on the wave vector anisotropy. Under the assumption of convected  turbulent fluctuations through the spacecraft  (i.e., the phase velocity $v_{\phi}$ of the fluctuations is small with respect to the flow velocity),  these measurements are possible with one spacecraft thanks to the variation of the mean magnetic field ${\bf B}$ direction  with respect to the bulk flow ${\bf V}$.  While ${\bf V \| B}$, the  spacecraft resolve fluctuations with ${\bf k \| B}$, when ${\bf V \perp B}$, the fluctuations with  ${\bf k \perp B}$ are measured.  

This idea was already used in the solar wind for studying the wave vector anisotropies of the Alfv\'enic fluctuations in the inertial range (Matthaeus et al., 1990; Bieber et al., 1996; Saur \& Bieber, 1999).  The authors suppose that the observed turbulence is a linear superposition of two uncorrelated components, slab and 2D, and both components have a power law energy distribution with the same spectral  index $s$, $\delta B_{\perp}^2(k_{\|})\sim A_1 k_{\|}^{-s}$ and  $\delta B_{\perp}^2(k_{\perp}) \sim A_2 k_{\perp}^{-s}$, where $A_1$ and $A_2$ are the amplitudes of slab and 2D turbulent components, respectively.  Bieber et al. (1996) propose two independent observational tests for distinguishing the slab component from the 2D component.  
 
The first test is based on the anisotropy of the power spectral density (PSD) of the magnetic fluctuations in the plane perpendicular to ${\bf B}$, i.e., on the non-gyrotropy of the PSD at a given frequency in the spacecraft frame: in the case of a slab turbulence with ${\bf k \| B}$, all the wave vectors suffer the same Doppler shift depending on the angle between ${\bf k}$ and ${\bf V}$,  and if  the spectral power is  gyrotropic in the plasma frame, it will remain gyrotropic in the spacecraft frame; in the case of a 2D turbulence, with ${\bf k \perp B}$, if the PSD is gyrotropic in the plasma frame, it will be non-gyrotropic in the spacecraft frame because the Dopler shift  will be different if ${\bf k}$ is perpendicular to ${\bf V}$ and if ${\bf k}$ has a component along ${\bf V}$.

The second test reveals the dependence  of the  PSD at a fixed frequency on the angle between the plasma flow and the mean field $\Theta_{BV}$ (defined between $0$ and $90$ degrees): For a PSD decreasing  with $k$ (like a power-law, for example),  in the case of the slab turbulence, the PSD for a given frequency will be more intense for $\Theta_{BV} = 0^{\circ}$, and therefore, the PSD decreases while $\Theta_{BV}$ increases;  for the 2D turbulence the PSD will be more intense for $\Theta_{BV} = 90^{\circ}$, and so it  increases  with $\Theta_{BV}$.  Using these tests, Bieber et al. (1996) have shown that the  inertial range of the slow solar wind is dominated by a 2D turbulence; however, a small percentage of a slab component is present. 

Mangeney et al. (2006) proposed a model of anisotropic wave vector distribution without any assumption on the independence of the two turbulence components. In their  gyrotropic model, the wave vector can be oblique with respect to  the $\|$ and $\perp$ directions.  The authors introduce a cone aperture of the  angle $\theta_{kB}$ between ${\bf k}$ and ${\bf B}$, as a free parameter of the model.  They assume  a power law  distribution of the total energy of the fluctuations $\sim k^{-s}$, with $s$ independent on $\theta_{kB}$.  For a given $k$, the turbulent spectrum is modeled by one of the two typical angular distribution $\sim  \cos(\theta_{kB})^{\mu}$ for ${\bf k}$ nearly parallel to ${\bf B}$ and   $\sim \sin(\theta_{kB})^{\mu}$ for ${\bf k}$ nearly perpendicular to ${\bf B}$.  For these two distributions,  the cone aperture of $\theta_{kB}$ is about $20^{\circ}$ for $\mu=10$ and $7^{\circ}$ for $\mu=100$.  The angle $\theta_{kB}$ can be easily represented through $\Theta_{BV}$ and so the model can be tested with one-spacecraft measurements.  

An advantage of the magnetosheath with respect  to the solar wind in ecliptical plane is that the angle $\Theta_{BV}$ covers the range from $0^{\circ}$ to $90^{\circ}$ within rather short time periods (one hour, or so) while other plasma conditions  remain roughly the same. 
A comparison of the model described above with the observations of the total PSD of the magnetic fluctuations within the magnetosheath flanks, at frequencies between $f_{lh}$ and $10 f_{ce}$, shows that these fluctuations have a strongly anisotropic distribution of  ${\bf k}$, with  $\theta_{kB}=(90 \pm 7)^{\circ}$ (Mangeney et al., 2006). Actually, this model (as well as  the tests of Bieber et al., 1996) is valid not only for a power law energy distribution in $k$, but for any monotone dependence where the energy decreases with increasing $k$.

Mangeney et al. (2006) have also shown that the variations of $\delta B^2$ with $\Theta_{BV}$ for a given frequency was not consistent with the presence of waves with a non-negligible phase velocity $v_\phi$. In other words,  if the observed turbulent fluctuations are a superposition of waves, their $v_\phi$ has to be much smaller than the flow velocity for any wave number $k$. This is consistent with the assumption that the wave frequency  $\omega_0$ is vanishing: the fluctuations are due to magnetic structures frozen in the plasma frame. These results have been obtained in the magnetosheath flanks for $f>10$~Hz, at electron  spatial scales $\sim[0.3,30] kc/\omega_{pe}$.

In this paper we extend the study of Mangeney et al. (2006) to  frequencies below 10 Hz, for the same time periods in the magnetosheath flanks. As a result, we will cover the largest possible scale range, from electron ($\sim 1$~km) to MHD scales ($\sim 10^5$~km).  At variance with the previous study, we analyse the spectral shapes and anisotropies for parallel ($\sim$~compressive) $\delta B_{\|}$ and for transverse ($\sim$~Alfv\'enic) $\delta B_{\perp}$ fluctuations independently.   For  Alfv\'enic fluctuations $\delta B_{\perp}$ we perform the first test of Bieber et al. (1996), i.e., we analyze the gyrotropy of  the PSD of the magnetic fluctuations in the plane perpendicular to  ${\bf B}$, at a given frequency, as a function of $\Theta_{BV}$.

 
\section{Data and methods of analysis}

 For our study we use high resolution  (22  vectors per second) magnetic field waveforms measured by  the FGM instrument  (Balogh et al., 2001).  Four seconds averages of the PSD of the magnetic fluctuations at 27 logarithmically spaced frequencies, between $8$~Hz and $4$~kHz, are measured by the STAFF Spectrum Analyser  (SA)  (Cornilleau-Wehrlin et al., 1997).  Plasma parameters with a time resolution of 4 seconds are determined from HIA/CIS measurements (R\`eme et al., 2001).

\subsection{Magnetic spectra and decomposition in $\delta B_{\perp}$ and $\delta B_{\|}$}

 High resolution FGM measurements allow to resolve turbulent spectra up to $\sim 10$~Hz.  Similar to Alexandrova et al. (2006), we calculate the spectra of the magnetic fluctuations in the GSE directions X, Y and Z,  using the  Morlet wavelet transform. The total power spectral density (PSD) is  $\delta B^2(f) = \sum_{j=X,Y,Z} \delta B^2_j(f)$.  The PSD of the compressive fluctuations $\delta B^2_{\parallel}(f)$ is  approximated by the PSD of  the modulus  of the magnetic field.  This is a good approximation when  $\delta B^2 \ll B_0^2$, where $B_0$ is the modulus of the magnetic field at the largest scale of the analysed data set.  The PSD of the transverse fluctuations is therefore 
\begin{equation}
\delta B^2_{\perp}(f) = \delta B^2(f) - \delta B^2_{\parallel}(f) .
\end{equation}
This approach, based on wavelet decomposition, allows the separation  of $\delta B_{\perp}$ and $\delta B_{\|}$  with respect to a local  mean field, i.e. to the field averaged on a  neighbouring scale  larger   than the scale of the fluctuations.  The lower frequency limit of this approach is a scale where  the ordering $\delta B^2 \ll B_0^2$ is no longer satisfied. 
   
The STAFF-SA instrument measures the spectral matrix $\langle \delta B_i(f)  \delta B_j(f) \rangle$ at higher frequencies.  Because of a recently detected  error  about the axes directions in the spin plane (O. Santolik, 2008, private communication) we cannot separate parallel and perpendicular spectra  at  the STAFF-SA frequencies; however we present here the total PSD, the trace of the spectral matrix.

\subsection{Anisotropy of the ${\bf k}$ distribution}

The motion of the plasma with respect to  a probe allows a 1D analysis of the wave vector distribution along the direction of  ${\bf V}$,  as was discussed in section 1.  The 3D wave vector  power spectrum $I({\bf k}) \equiv I(k,\theta_{kB},\varphi_k)$  depends on the wave number $k$, on the angle $\theta_{kB}$ between ${\bf k}$ and ${\bf B}$, and on the azimuth $\varphi_k$ of ${\bf k}$ in the plane perpendicular to ${\bf B}$. If $\omega_0$ is the frequency of a wave in the plasma rest frame ($\omega_0$ and $\omega$ are assumed to be positive), the Doppler shifted frequency $f = \omega /2 \pi$ in the spacecraft frame is given by 
\begin{equation}
\omega = |\omega_0 + {\bf k\cdot V}|.
\end{equation}
\noindent The trace of the power spectral density at this frequency is 
\begin{equation}
\delta B^2(\omega) = A \int  I({\bf k}) \delta(\omega -|\omega_0 + {\bf k\cdot V}|) d{\bf k} 
\end{equation}
\noindent i.e. the sum of the contributions with different ${\bf k}$.  $A$ is a normalisation factor and $\delta$ the Dirac function.

The angle $\theta_{kB}$  can be considered as depending on the angle  $\theta_{kV}$ between ${\bf k}$ and ${\bf V}$, the angle appearing in the Doppler shift frequency, and on the angle $\Theta_{BV}$ between ${\bf B}$ and ${\bf V}$ (see equation (2) of Mangeney et al., 2006).  Thus, the variations of $\delta B^2$ with $\Theta_{BV}$ for a given $\omega$ will give information about $I({\bf k})$. As was discussed in section~1,  $\delta B^2$ increases with increasing
$\Theta_{BV}$ when the fluctuations have $k_{\perp} \gg k_{\|}$ (2D turbulence) and it decreases for a slab turbulence  with $k_{\|} \gg k_{\perp}$ (see Figure 6 of Mangeney et al., 2006). As a consequence, in the case of 2D turbulence, the spectrum of the fluctuations will be higher for large angles $\Theta_{BV}$ than for small ones, and vice-versa for the slab geometry.

\subsection{$\delta {\bf B}$--anisotropy in the ${\bf BV}$--frame}

To study the distribution of the PSD $\delta B_{\perp}^2(f)$ in the plane perpendicular to ${\bf B}$, i.e. the gyrotropy of the magnetic fluctuations, taking into account the direction of the flow velocity ${\bf V}$, we  shall consider the following reference frame (${\bf b, bv, bbv}$):  ${\bf b}$ is the direction of the  ${\bf B}$ field, ${\bf bv}$ the direction of ${\bf B \times V}$ and ${\bf bbv}$ the direction of ${\bf B \times (B \times V)}$. 

The definition of this frame depends on  the considered scale (frequency). A local reference frame (defined on a neighbouring scale  larger   than the scale of the fluctuations)  can be defined only for frequencies below the 
spacecraft spin frequency $f_{spin}=0.25$~Hz which limits the plasma moments time resolution to $4$~s. That is why, for any frequency $f>f_{spin}$ we shall use the frame  (${\bf b, bv, bbv}$)  redefined every $4$~s. 

In this frame, we only consider the frequencies below $10$~Hz, i.e., the FGM data  (the STAFF-SA data cannot be used because  of the error  that has to be corrected in the whole data set). 
We project the wavelet transform of $B_X$,  $B_Y$ and $B_Z$ on the ${\bf b, bv, bbv}$
directions and we calculate the squares of these projections  $\delta B^2_b(f,t)$, $\delta B^2_{bv}(f,t)$ and $\delta B^2_{bbv}(f,t)$ which are the diagonal terms of the spectral matrix in this new frame.


\section{ ${\bf k}$-distribution of $\delta B_{\perp}$ and $\delta B_{\|}$}

\subsection{A case study with $\beta_p \simeq$ 1}
\begin{figure}[t]
  \begin{center}
   \includegraphics[width=8cm]{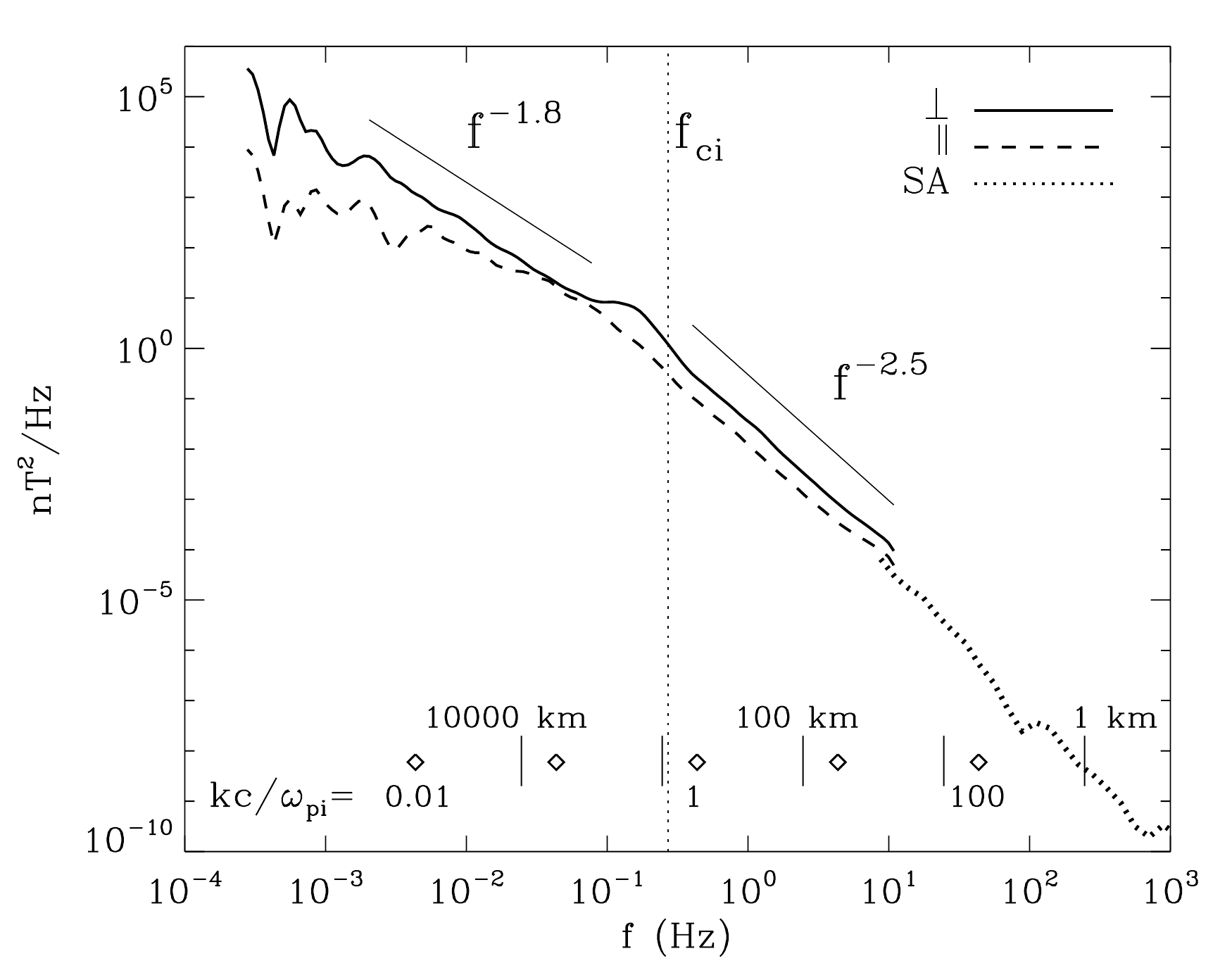}
  \end{center}
  \caption{\label{fig:fig1_I} FGM and STAFF-SA/Cluster data on 19/12/2001, 02:00-04:00~UT. Average spectra of the magnetic fluctuations, calculated using the Morlet wavelet transform of the FGM data ($f<10$~Hz). Solid line: for the transverse fluctuations $\delta B_{\perp}$. Dashed line: for the compressive fluctuations $\delta B_{\|}$.  Dotted line: the total power spectral density, STAFF-SA data ($f>8$~Hz). The diamonds give the scales $k c/\omega_{pi} \simeq k r_{gi} \simeq 0.01$ to $100$.  The vertical dotted line gives the average proton cyclotron frequency $f_{ci}$. The shapes of the power laws $f^{-1.8}$ and $f^{-2.5}$ are shown.}
\end{figure}

We consider an interval on the day 19/12/2001, from 02:00 to 04:00~UT.  
For this interval, the mean plasma parameters are the following:  the magnetic field $B=(18\pm 2)$~nT, the proton plasma density $N_p=(7 \pm 1)$~cm$^{-3}$, the proton temperature $T_p=(120\pm 15)$~eV, the proton plasma beta $\beta_p =(1.1\pm 0.4)$,  the ion  inertial  length $c/\omega_{pi}=(90\pm 5)$~km and the ion Larmor radius $r_{gi}=(65\pm 15)$~km. The average upstream bow shock angle $\theta_{BN}$ calculated with the ACE data is about $70^{\circ}$ (Lacombe et al., 2006). 

Figure~\ref{fig:fig1_I} displays the average spectra of the FGM data  for the transverse  fluctuations (solid lines) and for the compressive fluctuations (dashed lines).  The total PSD of the STAFF-SA data is the dotted line above $8$~Hz. The total covered frequency range is more  than six decades, from $3\cdot 10^{-4}$~Hz to $300$~Hz.  The small vertical bars just above the abscissae-axis indicate the scales from $\lambda = 10^4$~km to $1$~km corresponding to the Doppler shift  $f = V /  \lambda$ for  $\theta_{kV}=0^{\circ}$ and for the average velocity $V=(246\pm 25)$~km/s.  The diamonds above the abscissae indicate the scales $k c/\omega_{pi} \simeq k r_{gi} \simeq 0.01$ to $100$, corresponding to the frequency $f = k V / 2 \pi$.  Precisely,  $k c/\omega_{pi}=1$ appears in the spectrum at $f=(0.44\pm 0.05)$~Hz and  $k r_{gi}=1$ appears  at $f=(0.63\pm 0.11)$~Hz.

We see in Figure~\ref{fig:fig1_I} that, in the FGM frequency range, $\delta B^2_{\perp}(f)$ is everywhere larger than $\delta B^2_{\|}(f)$, except around $f\sim 5\cdot 10^{-2}$~Hz where $\delta B^2_{\perp} \sim \delta B^2_{\|}$, and where the compressive fluctuations display a spectral break. The spectrum of $\delta B_{\perp}$ displays a bump and a break around $0.2$~Hz, that can be a signature of Alfv\'en vortices \citep{olga06}. Below the bump, $\delta B^2_{\perp}(f) \sim f^{-1.8}$,  a power law with an exponent close to the Kolmogorov's  one  $-5/3$ (in section 5 we will analyse spectral shapes  in more details).  Above the bump, for $k c/\omega_{pi} > 0.2$, $\delta B^2_{\perp}(f)$ and $\delta B^2_{\|}(f)$ follow a similar power law $\sim f^{-2.5}$. This power law extends on the STAFF-SA frequency range up to  $k c/\omega_{pi} \simeq 50$ ($kc/\omega_{pe} \simeq 1.3$).    It is quite possible that, above these scales, the dissipation of the electromagnetic turbulence starts. However, around $f\simeq 100$~Hz, there is another spectral bump, which is due to whistler waves, identified by their right-handed polarisation. The question of the turbulence dissipation is out of scope of the present paper and will be analysed in details in the future.

\begin{figure}[t]
\begin{center}
\includegraphics[width=8cm]{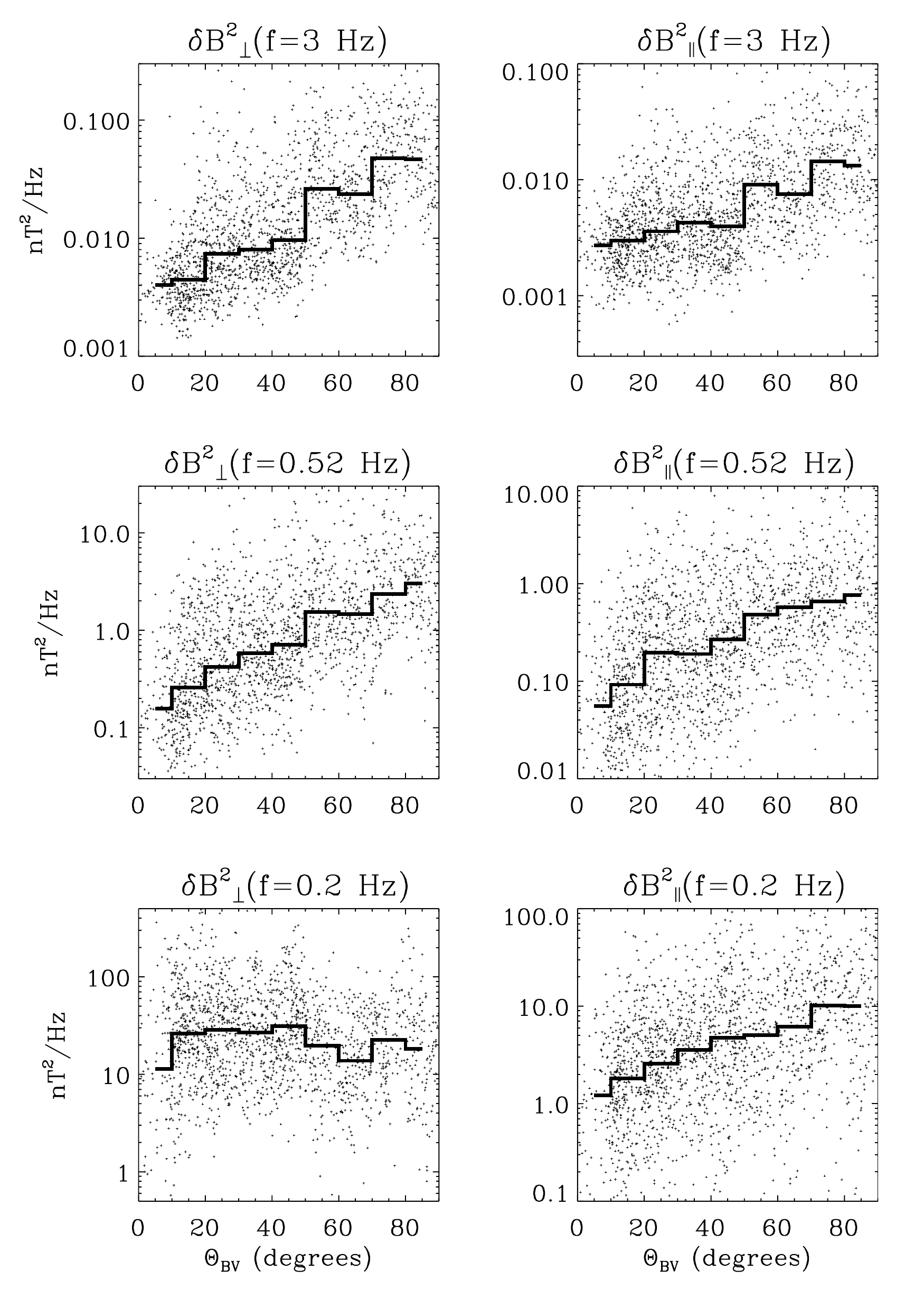}
\end{center}
\caption{FGM/Cluster data on 19/12/2001, 02:00-04:00~UT. Upper panels: Scatter plots of the power spectral density  of the magnetic fluctuations at $f=3$~Hz as a function of the angle between the local mean field and velocity, $\Theta_{BV}$. The distributions of the energy of Alfv\'enic fluctuations $\delta B_{\perp}^2$ is shown in the left panel, $\delta B^2_{\|}$ is shown in the right panel. Middle and lower panels have the same format, but here the frequencies are respectively $0.52$~Hz  and $0.2$~Hz. In all panels, the thick lines give the median value for bins $5^{\circ}$ wide.}
\label{fig:fig2}
\end{figure}

Now, we consider the anisotropy of the distribution of the  wave vectors.  Figure~\ref{fig:fig2} shows the dependence of $\delta B^2_{\perp}$ (left column) and $\delta B^2_{\|}$ (right column)  on the angle $\Theta_{BV}$ at different frequencies.  The thick solid curves give the median values for bins $5^\circ$ wide. The upper panels of Figure~\ref{fig:fig2} correspond to $f=3$~Hz ($kc/\omega_{pi} \simeq 7$).  The observed  increase of $\delta B^2_{\perp}$ and $\delta B^2_{\|}$ with $\Theta_{BV}$ can be produced only by fluctuations with $k_{\perp}\gg k_{\|}$,  with phase velocities $v_{\phi}$ negligible with respect to the plasma bulk velocity,  and with decreasing intensity of the fluctuations with increasing $k$ (as was discussed in sections 1 and 2.2).

At larger scales (lower frequencies) we observe the same tendency for $f>0.3$~Hz. The middle panels of Figure~\ref{fig:fig2} display $\delta B^2_{\perp}$ and $\delta B^2_{\parallel}$ as functions of $\Theta_{BV}$ for $f=0.52$~Hz ($kc/\omega_{pi} \simeq 1.2$):  we still observe here a clear increase of  $\delta B^2_{\perp}$ and $\delta B^2_{\parallel}$ with $\Theta_{BV}$. The lower panels of Figure~\ref{fig:fig2} display $\delta B^2_{\perp}$ and $\delta B^2_{\parallel}$ at $0.2$~Hz ($kc/\omega_{pi} \simeq 0.5$),  just at the spectral bump of $\delta B^2_{\perp}$ (see Figure~\ref{fig:fig1_I}). $\delta B^2_{\parallel}$ still increases with $\Theta_{BV}$  (in spite of a large dispersion of the data points around the  median), while  $\delta B^2_{\perp}$ has a flat distribution with $\Theta_{BV}$. This can be due to several reasons: (i) $I(k)$ is no longer  a decreasing function with $k$, (ii) $I(\theta_{kB})$ is more isotropic in the spectral bump and/or (iii) the fluctuations are not frozen in plasma at this scale. This spectral bump, as we have already mentioned, can be the signature of Alfv\'en vortices with $k_{\perp} \gg k_{\|}$, propagating slowly in the plasma frame. It can be also the signature of propagating AIC waves with $k_{\|} \gg k_{\perp}$, which are unstable for the observed plasma conditions (Mangeney et al., 2006; Samsonov et al., 2007). However,  as explained in sections~1 and 2.2, the energy of fluctuations with $k_{\|} \gg k_{\perp}$ would decrease with increasing $\Theta_{BV}$ at a given frequency,  while in our case $\delta B^2_{\perp}$ seems to be independent on $\Theta_{BV}$.  
 
As we have just seen from Figure~2 (and as discussed in sections~1 and~ 2.2), the comparison of the energy level of the turbulent fluctuations for different  $\Theta_{BV}$ at a given frequency gives us a good estimate of the wave vector anisotropy. We compare now the PSD of the fluctuations observed for large $\Theta_{BV}$ and for small $\Theta_{BV}$ in the whole frequency range, to estimate the wave vector anisotropy in a large domain of wave vectors. 

\begin{figure}[t]
  \begin{center}
  \includegraphics[width=8cm]{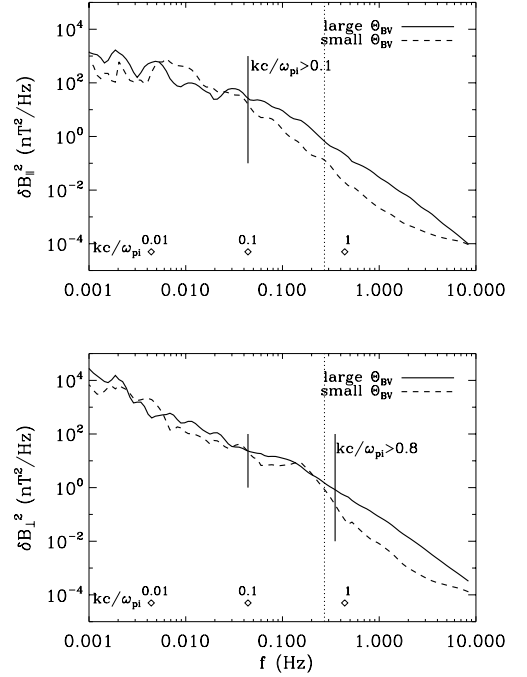}
  \end{center}
 \caption{\label{fig:fig4_I}
FGM/Cluster data on 19/12/2001, 02:00-04:00~UT.   Average power spectral density of the transverse magnetic fluctuations (upper panel) and of the compressive fluctuations (lower panel). In each panel, the solid line is the average spectrum for large $\Theta_{BV}$ angles, and the dashed line for small $\Theta_{BV}$. The vertical dotted line gives the average $f_{ci}$, the diamonds indicate $kc/\omega_{pi}=0.01$, $0.1$ and $1$.}
\end{figure}

The upper panel  of Figure~\ref{fig:fig4_I} displays the spectra of  the compressive fluctuations 
$\delta B^2_{\| la}$ for  the $10$\% of the points of the sample with the largest  $\Theta_{BV}$ ($la=$ large angles, solid line), and $\delta B^2_{\| sa}$ for the $10$\% of the points with the smallest $\Theta_{BV}$ ($sa=$ small angles, dashed line).   At frequencies below 0.06 Hz ($kc/\omega_{pi}=0.1$, indicated by a vertical solid line) the spectra  $\delta B^2_{\| sa} \simeq \delta B^2_{\| la}$.  At higher frequencies, $f>0.06$~Hz (i.e., at smaller scales, $kc/\omega_{pi} > 0.1$)  we observe $\delta B^2_{\| la} > \delta B^2_{\| sa}$.  This indicates that 2D turbulence dominates at such small scales.  
Close to $10$~Hz, i.e. at the vicinity of the FGM Nyquist frequency,  we see that  $\delta B^2_{\| la} \simeq \delta B^2_{\| sa}\simeq 10^{-4}$~nT$^2$/Hz, that is the sensitivity limit of the FGM instrument. Therefore, the observations at $f>5$~Hz are not physically reliable.

The lower panel of Figure~\ref{fig:fig4_I} displays the spectra for the transverse fluctuations for large and small angles $\Theta_{BV}$, $\delta B^2_{\perp la}$ (solid line) and  $\delta B^2_{\perp sa}$ (dashed line). We observe that  $\delta B^2_{\perp la}$  becomes larger than  $\delta B^2_{\perp sa}$ at about the same scale of $kc/\omega_{pi} \simeq 0.1$ as for compressive fluctuations. However, here within the spectral bump range, $\sim [0.1,0.3]$~Hz, we observe $\delta B^2_{\perp la} \simeq \delta B^2_{\perp sa}$.  This  is consistent with  our previous results that in this  short  frequency range the 2D turbulence model is not valid (cf. Figure~\ref{fig:fig2}).  A clear dominance of $\delta B^2_{\perp la}$ over $\delta B^2_{\perp sa}$  is  then observed for $f > 0.3$~Hz ($kc/\omega_{pi} > 0.8$, see a vertical solid line).


These observations allow to conclude  that, for $\beta_p \simeq 1$,  there is a change in the nature of the wave-vector distribution of the magnetic fluctuations in the magnetosheath, in the vicinity of ion characteristic scale: if at MHD scales there is no clear evidence for a dominance of a slab or 2D geometry of the fluctuations, at ion scales ($kc/\omega_{pi} > 0.1$) the 2D turbulence dominates. This is valid for both  the Alfv\'enic and compressive fluctuations.   The large scale  limit of the 2D turbulence is, however, different for Alfv\'enic and compressive fluctuations, and seems to depend on the presence of spectral features, as peaks or bumps. We analyse this point more in details by considering other cases.


\subsection{Other case studies}

The comparison between the spectra for large $\Theta_{BV}$ and for small $\Theta_{BV}$ has been made during four other one-hour intervals, with different average $\beta_p$ and different average shock angles $\theta_{BN}$. For the same intervals, Samsonov et al. (2007)  display the observed proton temperature anisotropy and the corresponding thresholds for AIC and mirror instabilities. 

\begin{figure}[t]
  \begin{center}
  \includegraphics[width=8cm]{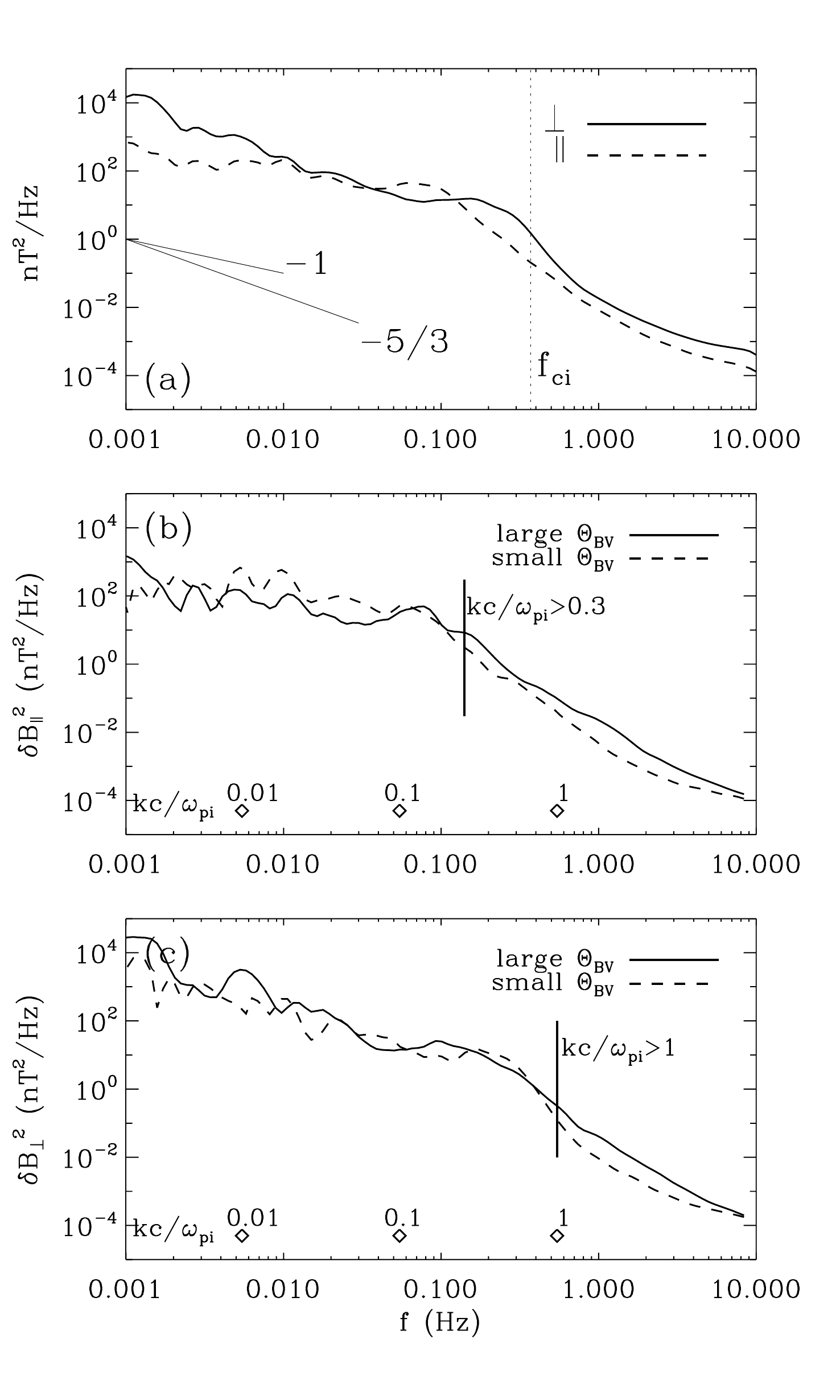}
  \end{center}
 \caption{\label{fig:fig5_I}
FGM/Cluster data on 17/05/2002, 08:30-09:30~UT. (a) Average PSD of $\delta B_{\perp}$ (solid line), PSD of $\delta B_{\|}$ (dashed line), the vertical dotted line gives the average $f_{ci}$; (b) the average spectrum of $\delta B_{\|}$ for large angles $\Theta_{BV}$ (solid line) and for small angles (dashed line), diamonds  indicate $kc/\omega_{pi}=0.01$, $0.1$ and $1$ ($kr_{gi} = kc/\omega_{pi}$ in this case); (c) same as (b), for $\delta B_{\perp}$. }
\end{figure}

Figure~\ref{fig:fig5_I}  gives the results of the analysis for an interval  (day 17/05/2002, 08:30-09:30~UT) for which $\theta_{BN} \simeq 70^{\circ}$,  $\beta_p = (1.6\pm 0.3)$,  $B=(24\pm 3)$~nT, $V=(190\pm 10)$~km/s, $f_{ci}=(0.37\pm 0.04)$~Hz, $c/\omega_{pi}=(55\pm 2)$~km and $r_{gi}=(50\pm 5)$~km. 

Figure~\ref{fig:fig5_I}a gives the average PSD of  transverse (solid line) and compressive  fluctuations  (dashed line).  There is a spectral bump for the transverse fluctuations around $0.2$~Hz. Below the spectral bump, $\delta B^2_{\perp} (f) \sim  f^{-5/3}$. For the compressive  fluctuations  there is a spectral bump around $0.07$~Hz, probably made of mirror modes. Below the bump, $\delta B^2_{\|} (f)$ is close to $f^{-1}$. 

In the two other panels of Figure~\ref{fig:fig5_I}, we display the spectra for large and small $\Theta_{BV}$   for compressive and for transverse fluctuations, respectively.    
In Figure~\ref{fig:fig5_I}b, at frequencies above the bump of $\delta B^2_{\|}$ ($f>0.1$~Hz, $k c/\omega_{pi} \geq$ 0.3) we observe  $\delta B^2_{\| la} > \delta B^2_{\| sa}$.  In Figure~\ref{fig:fig5_I}c, we observe  $\delta B^2_{\perp la} \simeq  \delta B^2_{\perp sa}$ at large scales  (observed at  $f < 0.05$~Hz, i.e. $kc/\omega_{pi}<0.1$), but  at frequencies above the bump of $\delta B^2_{\perp}$ ($k c/\omega_{pi} > 1$)  we observe $\delta B^2_{\perp la} >  \delta B^2_{\perp sa}$.  So, the transverse and compressive fluctuations have $k_{\perp} \gg k_{\|}$ at scales smaller than  their respective spectral bumps.  This confirms the conclusions of section 3.1.

For an interval  with  $\beta_p \simeq 0.8$ and $\theta_{BN} \simeq 70^{\circ}$  (day 16/12/2001, 08:00-09:00 UT) the average spectrum $\delta B^2_{\perp} (f)$ displays a spectral bump around 0.5 Hz. The comparison between the spectra for large and small $\Theta_{BV}$ (not shown) indicates that above the bump, for $k c/\omega_{pi} > 1$, the transverse fluctuations can be described by the 2D--turbulence model. For the compressive fluctuations, this model is valid for a larger range of scales, $k c/\omega_{pi} > 0.3$.  This confirms the results obtained for $\beta_p \simeq 1.6$, shown in Figure ~\ref{fig:fig5_I}  as well as the conclusions of section 3.1. 

 In an interval with a larger value of $\beta_p$ (day 17/05/2002, 11:00-12:00~UT, $\beta_p \simeq$ 4.5, $\theta_{BN} \simeq 73^{\circ}$), the analysis of the spectra for large and small $\Theta_{BV}$ (not shown) shows that the 2D turbulence takes place for $k c/\omega_{pi} > 0.3$ for the transverse fluctuations, and for $k c/\omega_{pi} > 0.2$ for the compressive fluctuations. So, the 2D turbulence range of scales  for the transverse fluctuations is wider in this case.

\begin{figure}[t]
  \begin{center}
  \includegraphics[width=8cm]{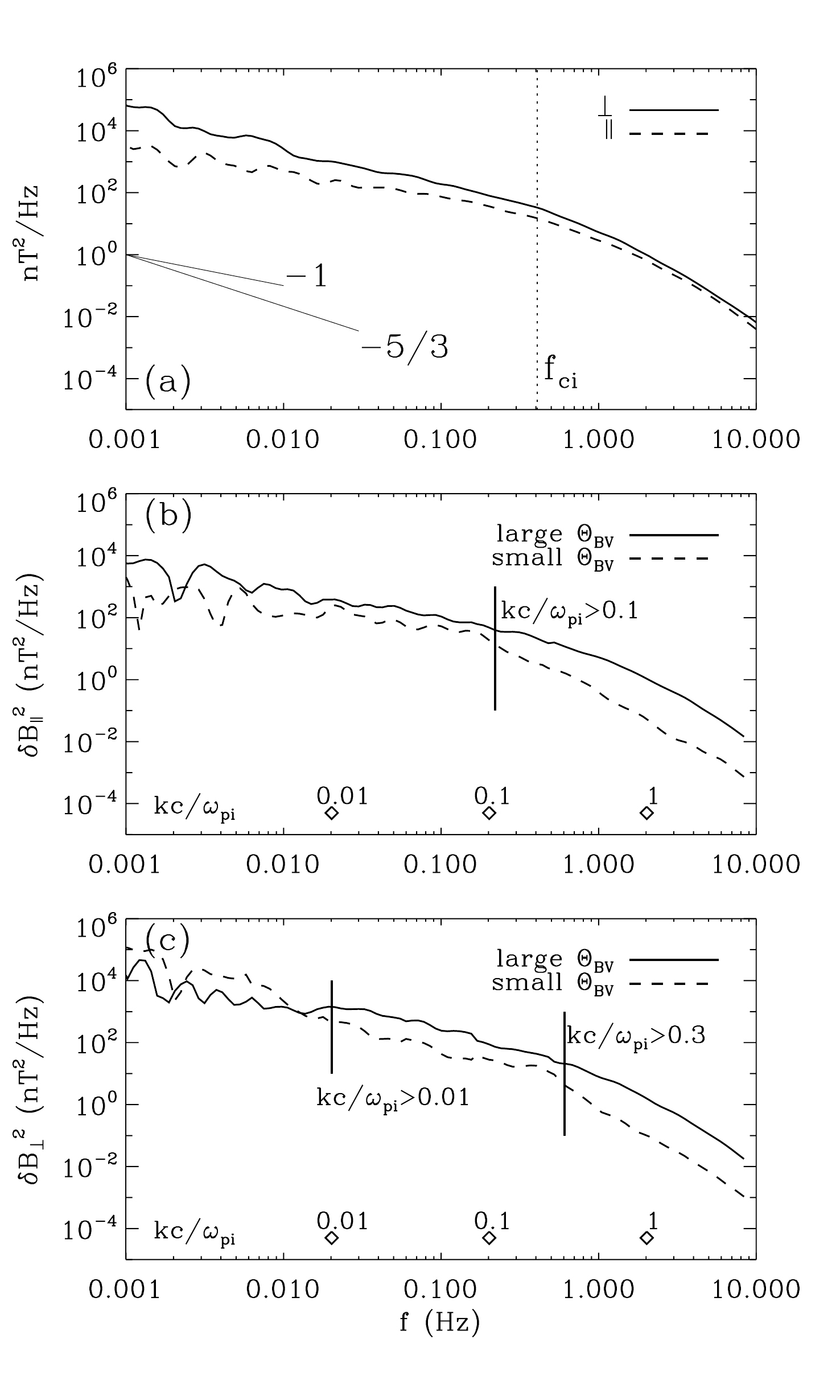}
    \end{center}
 \caption{\label{fig:fig6_I}
Same as Figure~\ref{fig:fig5_I}, but for FGM/Cluster data on 16/12/2001, 05:30-06:30~UT.}
\end{figure}

Figure~\ref{fig:fig6_I} shows the results of the analysis  for an interval   (day 16/12/2001, 05:30-06:30~UT) downstream of an oblique bow shock ($\theta_{BN} \simeq 50^{\circ}$), when $\beta_p=(9\pm3)$, $B=(27\pm 6)$~nT, $V=(370\pm20)$~km/s, $f_{ci}=(0.4\pm0.1)$~Hz, $c/\omega_{pi}=(30 \pm 2)$~km and $r_{gi}=(60 \pm 20)$~km. The wavenumber $kc/\omega_{pi}=1$ appears  in the spectrum at $f=(2.0 \pm  0.2)$~Hz and the 
 wavenumber $kr_{gi}=1$ appears at $f=(1.1 \pm 0.3)$~Hz. 

Figure~5b shows that at low frequencies (i.e., at large scales,  $k c/\omega_{pi} < 0.1$) the spectrum  for large angle  $\delta B_{\| la}$ dominates slightly at every frequencies. At smaller scales,  $k c/\omega_{pi} > 0.1$,  this dominance is more clear. Figure~5c shows that $\delta B_{\perp la} > \delta B_{\perp sa}$ as far as $k c/\omega_{pi} > 0.01$ (wavelengths smaller than $10^4$~km), i.e.,  the $\delta B_{\perp}$ fluctuations can be described by the 2D--turbulence model at all the scales smaller than the Earth's radius.   In this case, with large value of plasma beta, the 2D turbulence range of scales increases again, but  the lower limit of 2D turbulence is not related to any spectral features, as was observed for smaller $\beta_p$. 

We may therefore conclude that, at frequencies above the spectral break in the vicinity of $f_{ci}$,  the $\delta B_{\perp}$ and $\delta B_{\|}$ fluctuations in the magnetosheath  have ${\bf k}$ mostly perpendicular to ${\bf B}$, and this is independent on $\beta_p$ and on  $\theta_{BN}$. In terms of  spatial scales,  this is valid  for  $kc/\omega_{pi} >0.1$ (or $>1$ when spectral features  appear in the vicinity of $kc/\omega_{pi}=1$). For high values of $\beta_p$, the range of scales of  the 2D turbulence seems to increase: for $\beta_p\sim 10$ the fluctuations have $k_{\perp} \gg k_{\|}$ for $kc/\omega_{pi} >0.01$.  This small scale spectral anisotropy is also independent on the presence of transverse and/or compressive spectral features (peaks) at larger scales.  Nevertheless, for the moderate values of beta ($\beta_p<3$), these spectral peaks appear as the lower limit of 2D turbulence.


\section{Gyrotropy of the magnetic fluctuations}

In this section we  analyse the anisotropy of the amplitudes of magnetic fluctuations in the plane perpendicular to ${\bf B}$. For this purpose, we use the coordinate frame based on ${\bf B}$ and ${\bf V}$, ${\bf (b,bv,bbv)}$, as explained in section~2.3.

\begin{figure}[t]
  \begin{center}
  \includegraphics[width=8cm]{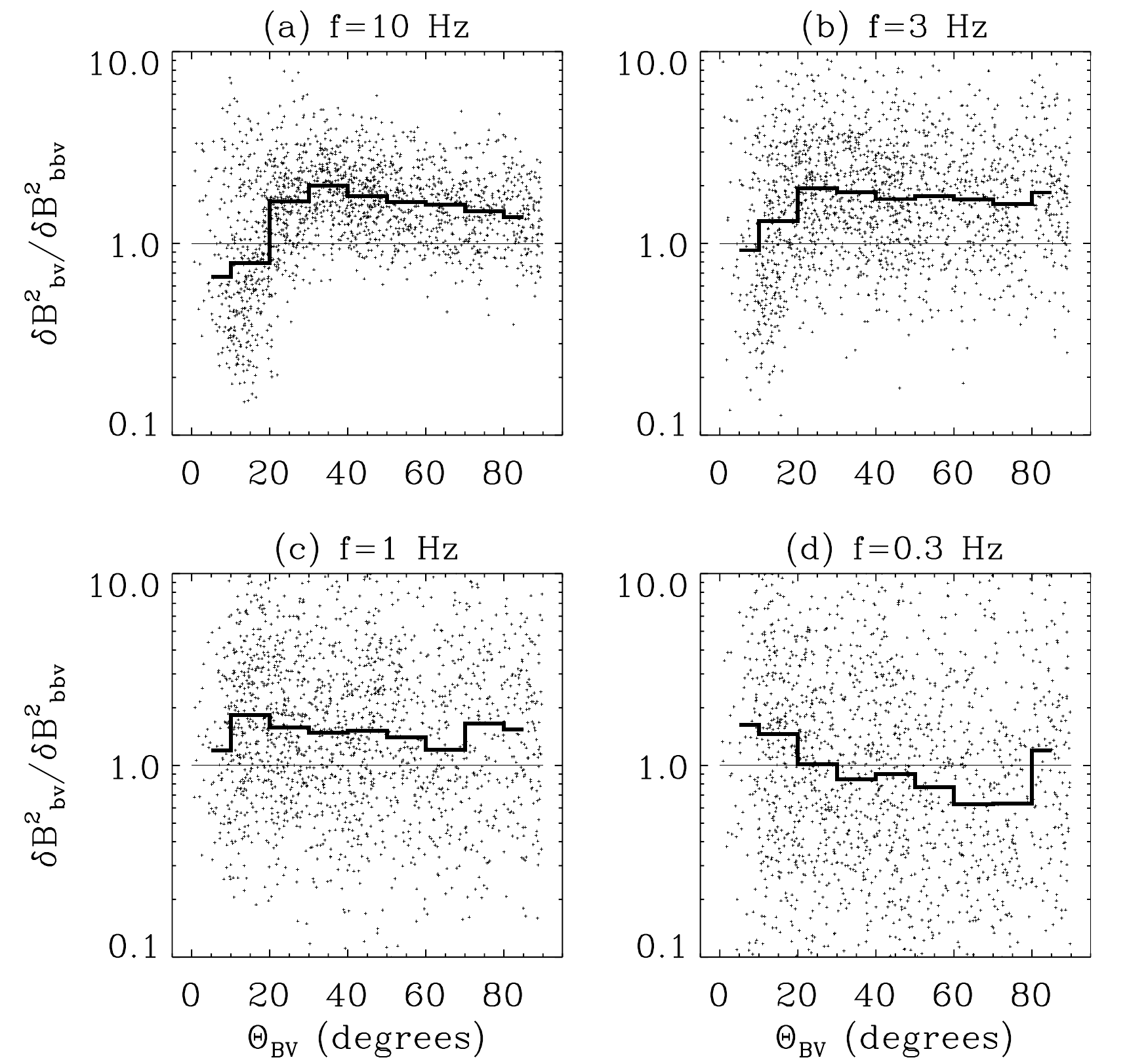}
  \end{center}
 \caption{\label{fig:fig7_I}
FGM/Cluster data on 19/12/2001, 02:00-04:00~UT. Scatter plots of the ratio $R=\delta B_{bv}^2$/$\delta B_{bbv}^2$ averaged over 4 s, as a function of the angle $\Theta_{BV}$ at (a) $10$~Hz, (b) $3$~Hz, (c) $1$~Hz and (d) $0.3$~Hz. The thick lines give the median value for bins $10^{\circ}$ wide.
}
\end{figure}

For the same time interval as Figure~\ref{fig:fig1_I},  Figure~\ref{fig:fig7_I}  displays the ratio $R=\delta B_{bv}^2$/$\delta B_{bbv}^2$, amplitude of the fluctuations along ${\bf B\times V}$ over the amplitude along  ${\bf B\times (B\times V)}$, in the plane perpendicular to ${\bf B}$, at four fixed frequencies, as a function of  $\Theta_{BV}$.

In Figure 6a ($f=10$~Hz, $k c/\omega_{pi} = 23$) and Figure 6b ($f=3$~Hz, $k c/\omega_{pi} = 7$) the ratio $R$ is larger than 1 for $\Theta_{BV} \gtrsim 20^{\circ}$. The median value decreases and reaches $1$ or less for $\Theta_{BV} < 20^{\circ}$. A similar dependence was observed by Bieber et al. (1996)  and Saur \& Bieber (1999) at MHD scales in the solar wind, indicating the dominance of the 2D turbulence.

   At larger scales (Figure 6c, $f=1$~Hz, $k c/\omega_{pi} = 2$),  in spite of the strong dispersion of $R$, the median  values are slightly  larger than $1$ for any $\Theta_{BV}$: the 2D turbulence still dominates.  At  an even larger scale, the scale of the spectral break  (Figure 6d, $f=0.3$~Hz, $k c/\omega_{pi} = 0.7$), the ratio $R$ is strongly dispersed.  The variation of the median  does not correspond to  a slab  or 2D turbulence. That is in agreement with the results obtained  within the spectral break frequency range in section 3.1 (cf. Figure 2, lowest panel for $\delta B_{\perp}$).

\begin{figure}[t]
  \begin{center}
 \includegraphics[width=8cm]{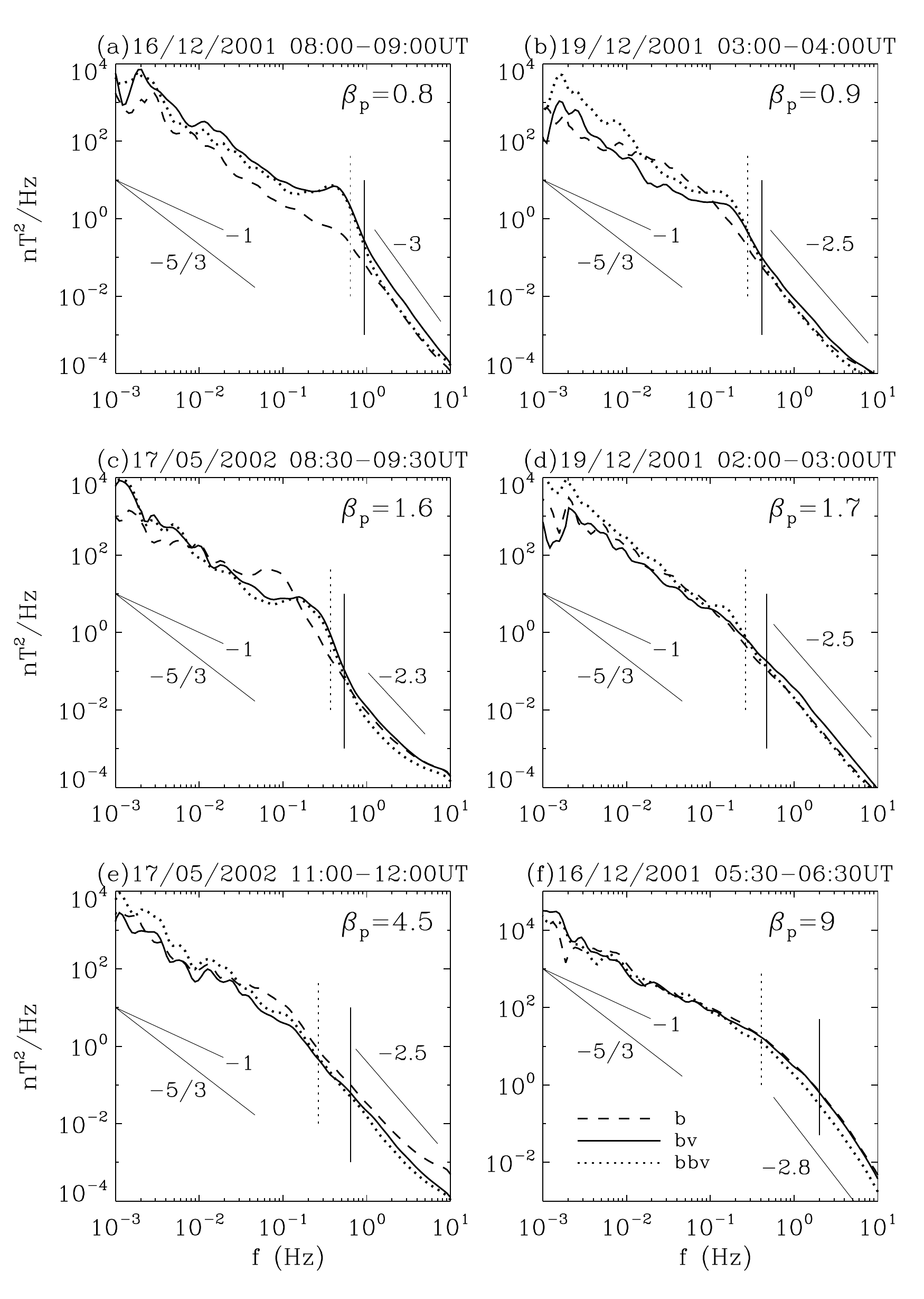}
  \end{center}
 \caption{\label{fig:fig8_I}
Average spectra of the magnetic fluctuations in the ${\bf (b,bv,bbv)}$--frame,
 ${\bf b}$ is parallel to the ${\bf B}$ field (dashed line), ${\bf bv}$ is parallel to ${\bf B \times V}$ 
(solid line), ${\bf bbv}$ is parallel to ${\bf B \times (B \times V)}$ (dotted line). For each of the 6 considered one-hour intervals  a vertical dotted bar gives $f_{ci}$, a vertical solid bar gives the Doppler shifted wavenumber $k c /\omega_{pi}=1$. In each panel the shapes of the power laws $f^{-5/3}$,  $f^{-1}$ are indicated;  in the high frequency range we show the $f^{-s}$ spectral shape, with $s$ determined in section 5, see Figure~8. }
\end{figure}

The anisotropy of the magnetic fluctuations  for the $[10^{-3},10]$~Hz frequency range is shown in Figure~\ref{fig:fig8_I} with average spectra in the three directions ${\bf  b}$ (dashed line), ${\bf  bv}$ (solid line) and ${\bf  bbv}$ (dotted line). The panels (a) to (f) correspond to increasing values of $\beta_p$ for the six considered intervals. For each interval, the vertical solid bar indicates the scale $k c/\omega_{pi} = 1$ and  the dotted bar shows $f_{ci}$. 

In Figures~\ref{fig:fig8_I}a, b and d,  for  $k c/\omega_{pi} \geq 1$ we observe that  the spectra of the components along ${\bf b}$ and along ${\bf bbv}$  are nearly equal, $\delta B_{b}^2 \simeq \delta B_{bbv}^2$. In Figures~\ref{fig:fig8_I}e and f, $\delta B_{b}^2$ is larger than $\delta B^2_{bv}$: the fluctuations are more compressive for the largest values of $\beta_p$.  All the panels of Figure~\ref{fig:fig8_I} show that $\delta B^2_{bv}> \delta B^2_{bbv}$  for $k c/\omega_{pi} \geq 1$. So, within the 2D turbulence range the PSD is not gyrotropic at a given frequency.  


As we  have mentioned  in section 1, the observed non-gyrotropy  in the spacecraft frame can be due 
to the Doppler shift. Indeed, we have shown in section~3 that the wave vectors ${\bf k}$ are mainly perpendicular to ${\bf B}$,  i.e ${\bf k}$  lies in plane spanned by  ${\bf bv}$ and ${\bf bbv}$. 
Assuming plane 2D turbulence, the relation ${\bf k_{\perp}\cdot \delta B}=0$ holds and thus, the wave vectors along ${\bf bbv}$ (i.e., along the direction of the flow in the plane perpendicular to ${\bf B}$, we denote such wave vectors ${\bf k_{bbv}}$) contribute to the PSD of  $\delta B_{bv}$ and the wave vectors along ${\bf bv}$ (${\bf k_{bv}}$) contribute to the PSD of $\delta B_{bbv}$.  Even if $I({\bf k})$ is gyrotropic, the fluctuations  $\delta B_{bv}$ with ${\bf k_{bbv}}$  suffer a  Doppler shift stronger than the fluctuations  $\delta B_{bbv}$  with ${\bf k_{bv}}$. For 2D turbulence, this Doppler shift effect is more pronounced  when $\Theta_{BV}$ reaches $90^{\circ}$.

This implies that, for a gyrotropic energy distribution, $I(k_{bv}) \simeq I(k_{bbv})$, in the plane perpendicular to ${\bf B}$, and  if the energy decreases with $k$, for example as a power law $I(k) \sim k^{-s}$,  the observed frequency spectrum  $\delta B^2_{bv}(f)$ will be more intense than $\delta B^2_{bbv}(f)$. In other words, at the same frequency $f$ in the spacecraft frame, we observe the fluctuations with $|{\bf k_{bv}}|>|{\bf k_{bbv}}|$. As the larger wave numbers correspond to a weaker intensity (for a monotone energy decrease with $k$), $\delta B_{bbv}^2$ will be smaller than $\delta B_{bv}^2$. 

 Therefore, the non-gyrotropy of $\delta B^2(f)$, observed here,  could be due to the non-gyrotropy of the Doppler shift, and could be compatible with a gyrotropic distribution of $I({\bf k})$. This is confirmed by the upper panels of Figure~\ref{fig:fig7_I}: as far as ${\bf k}$  is mainly perpendicular to ${\bf B}$,  the Doppler shift is small and gyrotropic for small $\Theta_{BV}$ and we observe  $R\sim 1$, i.e. the PSD is gyrotropic; but  for large  $\Theta_{BV}$, $R >1$.

On the other hand, the ratio $R(f)>1$, observed in Figures 6a and 6b for $\Theta_{BV} \simeq 90^{\circ}$, is also compatible with the non-gyrotropic ${\bf k}$--distribution observed by Sahraoui et al. (2006)  near the magnetopause, for $\Theta_{BV} \simeq 90^{\circ}$. In this case study, the authors show that  the turbulent cascade develops  along ${\bf V}$, perpendicular to ${\bf B}$ and ${\bf n}$, where ${\bf n}$ is  the normal to the magnetopause.  In this geometry, the direction ${\bf V}$ is close  to ${\bf bbv}$. Therefore, a non-gyrotropic $I({\bf k})$ distribution with  $\delta B^2_{bv}(k) / \delta B^2_{bbv}(k) > 1$  is expected in the ${\bf k}$--domain.  This non-gyrotropy of wave vectors is then reinforced by the Doppler shift, and would give $\delta B^2_{bv}(f) / \delta B^2_{bbv}(f) > 1$ in the $f$-domain.


 \section{Spectral shapes}

\begin{figure}[t]
  \begin{center}
 \includegraphics[width=8cm]{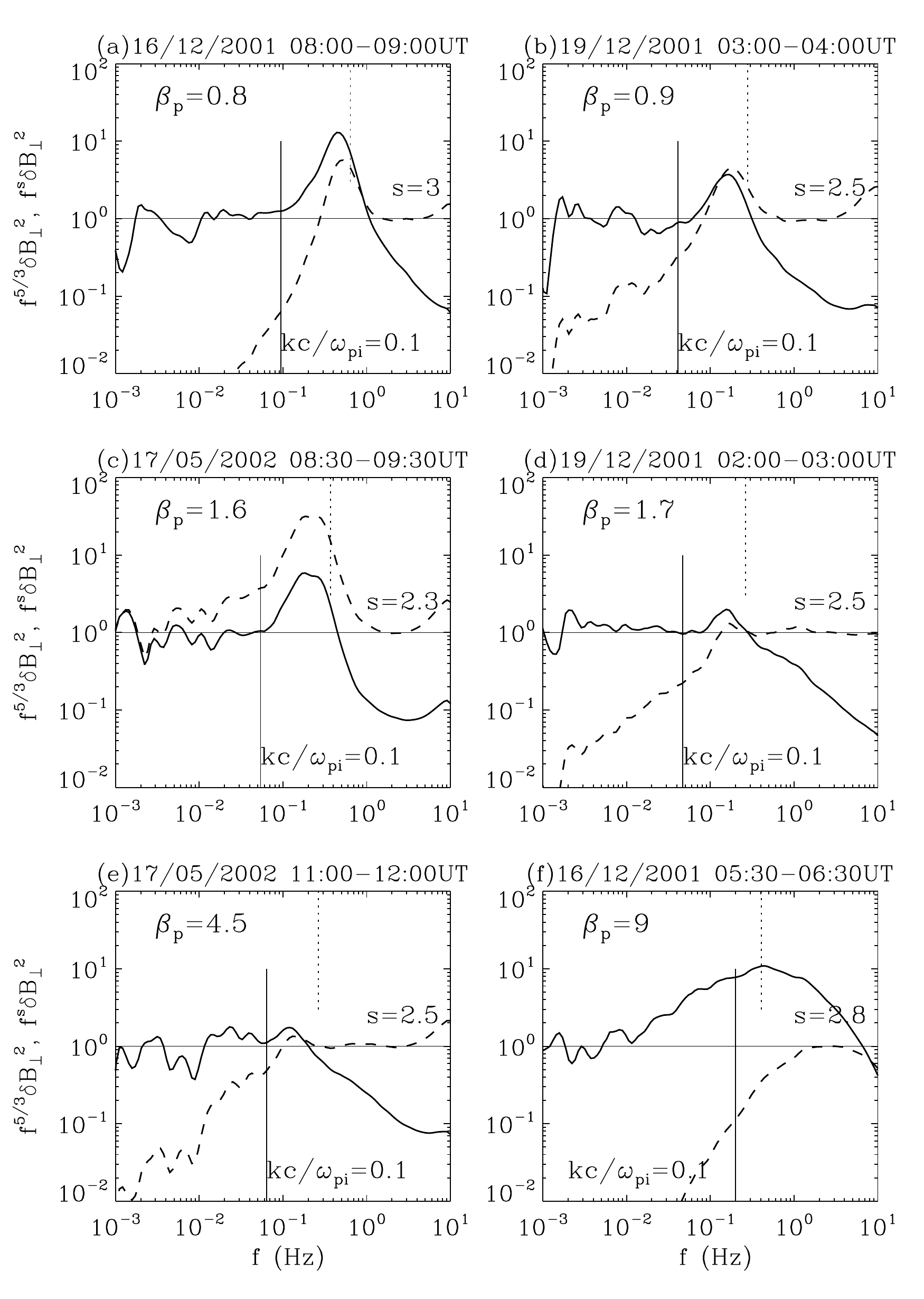}
  \end{center}
 \caption{\label{fig:fig9_I}
Compensated spectra $ f^{5/3} \delta B^2_{\perp}(f) $ (solid lines) and   $ f^{s} \delta B^2_{\perp}(f) $ with $s$ indicated in each panel (dashed lines) for the same time periods as Figure 7. }
\end{figure}

We have mentionned in section 3 that, at frequencies  $f<f_{ci}$, below the spectral break,  the spectra of the  transverse fluctuations $\delta B^2_{\perp}(f)$  follow a power law close to $f^{-5/3}$ (see Figure 1 and 4a). For the six intervals of Figure~7, Figure~8 displays compensated plots  of the transverse spectra $f^{5/3} \delta B^2_{\perp}(f)$ (solid lines). On the low frequency side of these plots, we see that the compensated spectra oscillate around a horizontal line, in a frequency range which varies slightly from day to day: a power law $ f^{-5/3}$ is thus a good  approximation for the observations in this frequency range. In Figures 8a to 8e, for which  $\beta_p$ is between 0.8 and 4.5, the Kolmogoroff $ f^{-5/3}$ power law is observed below 0.06 or 0.1 Hz, corresponding to scales  $kc/\omega_{pi }< 0.1$ (see the vertical solid bar).

In Figure 8f, for $\beta_p \simeq (9 \pm 3)$, the $ f^{-5/3}$ power law is only observed below $0.01$~Hz, i.e., below the 2D turbulence range of $\delta B_{\perp}$ (see Figure~5c). Above this frequency, as we see in  Figure 7f, the spectra of all the components are close to an $f^{-1}$ power law and the three spectra  have nearly the same intensity, $\delta B^2_{b}(f)\simeq \delta B^2_{bv}(f)\simeq \delta B^2_{bbv}(f) $.
This isotropy of the amplitudes of the turbulent fluctuations is natural to observe in  a high beta plasma, where the mean magnetic field  does not play any important role. Within this frequency range the spectrum can be also formed by a superposition of AIC and mirror waves. For such high $\beta_p$, the mirror modes are more unstable than the AIC waves, but they have an important Alfv\'enic component (G\'enot et al., 2001),  that can also contribute to make fluctuations more isotropic.

Above $f_{ci}$, the spectra $\delta B^2_{\|}(f)$ and $\delta B^2_{\perp}(f)$ follow similar power laws, see Figure~7.  The compensated spectra $f^{s} \delta B^2_{\perp}(f)$ with $s$ between 2 and 3 are presented in Figure 8 by dashed lines. We see that the composed spectra $f^{2.5} \delta B^2_{\perp}(f)$  oscillate around a horizontal line in a few cases (Figures 7b, 7d and 7e), for different values of $\beta_p$. In Figure 8a, the power law is steeper, $s=3$. Actually, in this case the spectral bump is the most clearly pronounced of the six analyzed intervals.  This bump is a signature of the Alfv\'en vortices, which have their own spectrum $\sim k^{-4}$ or $k^{-6}$, depending on the vortex topology (Alexandrova, 2008).  The superposition of the background turbulence with the coherent structures, like magnetic vortices, can  produce the observed  steep spectrum. In Figure 8c we do not observe any clear evidence for a power law spectrum in this frequency range.

\section{Summary and Discussion}

In the present paper, we have analysed six one-hour intervals in the middle of the terrestrial magnetosheath (at more than one hour from the crossing of the bow shock or of the magnetopause).  Precisely, we considered intervals in the magnetosheath flanks: the local times for the three considered days are respectively 8, 17 and 18~h (Lacombe et al., 2006). The proton beta varies from one interval to another, $\beta_p \in [0.8,9]$, that allows us to study the plasma turbulence in a very large range of plasma conditions.

\subsection{Spectral shape}

The spectral shape of the magnetic fluctuations in the magnetosheath has been studied by several authors (see Alexandrova, 2008). Rezeau et al. (1999) find a power law $f^{-3.4}$ above $f_{ci}$ in an interval close to the magnetopause. For intervals close to the bow shock, Czaykowska et al. (2001) find power laws around $f^{-1}$ below $f_{ci}$, and $f^{-2.6}$ above $f_{ci}$.  But in these studies,  intervals of $4$~minutes have been analyzed, so the minimal resolved frequency is about  $10^{-2}$~Hz.  In the present paper, the length of the intervals allows to reach frequencies smaller than $10^{-3}$~Hz. 

 Here, we present, for the first time, the observations of a Kolmogorov-like inertial range for Alfv\'enic fluctuations $\delta B^{2}_{\perp}(f)\sim f^{-5/3}$ in the frequency range $f<f_{ci}$. It is  clearly observed in  five of the six studied intervals, those for which $\beta_p < 5$ and when Alfv\'enic fluctuations were dominant. Such a Kolmogorov power law is observed in the Alfv\'enic inertial range of the solar wind turbulence, below the spectral break in the vicinity of $f_{ci}$. The presence of such power law in the magnetosheath flanks is consistent with the estimations made  by Alexandrova (2008): in the flanks, the transit time of the plasma is longer than in the subsolar regions, and it is much longer than the time of nonlinear interactions;  therefore, the turbulence has enough time to become developed.

In the high frequency range, $f>f_{ci}$, we generally observe  $\delta B_{\perp}^2(f)$ and $\delta B_{\|}^2(f)$ following similar power law $f^{-s}$ with   a spectral index  $s$ between $2$ and $3$, in agreement with previous studies.

\subsection{Wave-vector anisotropy}

We analysed here the anisotropy of  wave-vector distribution of the magnetic fluctuations 
 from $10^{-3}$  to $10$~Hz. This frequency range corresponds to the spatial scales going from $\sim 10$ to $10^5$~km (from electron to MHD scales). For this analysis we used a statistical method, based on the dependence of the observed  magnetic energy at a given frequency on the Doppler shift for different wave vectors \citep{bieber,Horbury2005,mangeney06}. 

Within the inertial range of the magnetosheath turbulence ($f<f_{ci}$, $k c/\omega_{pi} < 1$, $k r_{gi} < 1$), we do not observe a clear evidence of wave-vector anisotropy. It can be related to the fact that linearly unstable modes, such as AIC modes with ${\bf k}$ mainly parallel to ${\bf B}$  and mirror modes with ${\bf k}$ mainly perpendicular to ${\bf B}$, together with Alfv\'en vortices  with $k_{\perp}\gg k_{\|}$ co-exist in this frequency range.

However,  above the spectral break in the vicinity of the ion characteristic scales ($f>f_{ci}$, $k c/\omega_{pi} > 1$, $k r_{gi} > 1$ and up to electron scales), 
we observe a clear evidence of 2D turbulence with  $k_{\perp}\gg k_{\|}$ for both $\delta B_{\perp}$ and $\delta B_{\|}$, and independently on  $\beta_p$, on the  bow-shock geometry $\theta_{BN}$, and on the  wave activity within the inertial range at larger scales. This wave vector anisotropy  seems  to be a  general property of the small scale turbulence in the Earth's magnetosheath.

The range of wavenumbers of this 2D turbulence sometimes goes down to $k c/\omega_{pi} \simeq 0.1$ (or even to  $k c/\omega_{pi} \simeq 0.01$), but usually it is limited by  $k c/\omega_{pi} \simeq 1$, while at $k c/\omega_{pi} < 1$ spectral features (peaks or bumps) appear. As we can conclude from  the work of Mangeney et al. (2006), the  largest wavenumbers of the 2D turbulence are observed around $k c/\omega_{pi} \sim 100$, where the dissipation of electromagnetic turbulence begins. This last conjecture must be verified by a deeper analysis. 

\subsection{Anisotropy of magnetic fluctuations}

Analyzing the anisotropy of the amplitudes of turbulent fluctuations, we usually find that $\delta B^2_{\perp} > \delta B^2_{\|}$; but  for the largest plasma $\beta_p$, the fluctuations are more isotropic $\delta B^2_{\perp} \sim  \delta B^2_{\|}$. This is valid for both the large  scale inertial range and  the small scale 2D turbulence. The dominance of $\delta B^2_{\|}$ happens only locally in the turbulent spectrum, indicating the presence of an unstable  mirror mode.

Concerning the gyrotropy of the  amplitude of the magnetic fluctuations in the plane perpendicular to ${\bf B}$, there is no universal behavior at large scales.   At smaller scales, within the frequency range $[0.3-10]$~Hz and for any $\beta_p$, the  2D turbulence is observed to be non-gyrotropic: the energy   $\delta B^2_{bv}$  along the  direction perpendicular to ${\bf V}$ and ${\bf B}$ is larger than the energy $\delta B^2_{bbv}$ along the projection of ${\bf V}$ in the plane perpendicular to  ${\bf B}$.   This non-gyrotropy  might be a consequence of different Doppler shifts for fluctuations with ${\bf k}$ parallel and perpendicular to ${\bf V}$ in the plane perpendicular to ${\bf B}$. The non-gyrotropy at a fixed $f$ is compatible with  gyrotropic fluctuations at a given ${\bf k}$.   On the other hand such  a non-gyrotropy will be also observed if the ${\bf k}$--distribution is not gyrotropic, but is aligned with the plasma flow, as was observed by Sahraoui et al (2006) in the vicinity of the magnetopause.

\begin{acknowledgements}
We thank Jean-Michel Bosqued for providing the CIS/HIA proton data,  and Nicole Cornilleau-Wehrlin for the STAFF-SA data. We thank Joachim Saur for constructive comments on the paper.  We are very grateful to the team of the Cluster Magnetic field investigation, to the team of the STAFF instrument, and to the Cluster Active Archive (CAA/ESA).  
\end{acknowledgements}


\end{document}